# Helically-trapped electron mode in optimized stellarator

Javier H. Nicolau[1,*,&], Xishuo Wei[1,&], Pengfei Liu[1], Gyungjin Choi[1,$], and Zhihong Lin[1,*]

[1] Department of Physics and Astronomy, University of California, Irvine, CA 92697, USA

Global gyrokinetic simulations find a strong helically-trapped electron mode (HTEM) driven by density gradients in the optimized stellarator W7-X fusion experiment. The eigenmode structure localizes in the inner side of the torus with an unfavorable magnetic curvature and weak magnetic field, where there is a large fraction of helically-trapped electrons. The mode is excited mostly by the ion free energy and propagates in the ion direction with a linear frequency much smaller than the diamagnetic frequency. The instability saturates by nonlinear processes of turbulence spreading in the real space and spectral transfer from unstable to damped regions. The steady state HTEM turbulence drives a large particle flux that may have significant implications for the confinement of fusion fuel and removal of fusion ash in the stellarator reactor.

[*]E-mail: javier.hn@uci.edu; zhihongl@uci.edu
[&]These authors contributed equally to this work
[$]Present address: Seoul National University, Korea

Stellarator with non-axisymmetric magnetic configuration [1-3] has attracted intense research interest as a promising fusion reactor with the steady state operation and reduced risk of disruptions since no plasma current drive is needed. Advances in stellarator design have drastically reduced the collisional (so-called neoclassical) transport in the optimized stellarators such as the quasi-isodynamic W7-X stellarator [4,5]. With neoclassical transport reduced, microturbulence driven by driftwave instability [6] can dominate the particle and heat transport in the W7-X experiments as shown by power balance analysis [7,8]. Measurements of density fluctuations also exhibit characteristics of ion temperature gradient (ITG) and trapped electron mode (TEM) turbulence [9].

While the ITG turbulence can drive large heat flux, the TEM turbulence can drive large particle flux. The TEM instability excited by density gradients has been predicted theoretically to be stable in the quasi-isodynamic stellarators with favorable bounce-averaged magnetic curvature [10]. Consistently, linear local (flux-tube) gyrokinetic simulations find only weak TEM with mode amplitude peaks in the outboard midplane of the W7-X [11] and driven mostly by ions [12]. However, flux-tube simulation results could depend on the choice of the magnetic field line being simulated and are in general different from the full flux-surface simulations for the driftwave instability and zonal flow dynamics in the W7-X [13], because different field lines are not equivalent in the non-axisymmetric stellarator. Therefore, global simulation is required to study the microturbulence in the stellarators.

Global gyrokinetic simulations of the stellarators [14-16] has recently been performed using kinetic electrons, but remain a computational grand challenge due to the large ion-to-electron mass ratio and huge scale separation. Reduced mass ratio and spatial resolution are often used in nonlinear simulations [17]. In the current work, we utilize the unique capability of the global gyrokinetic GTC [18] for the first global simulation of the electrostatic microturbulence in the W7-X using real mass ratio and sufficient spatial resolution. Such a global simulation with unprecedented realism and resolution only becomes feasible thanks to GTC's efficient method of solving the electron drift kinetic equation [19-21], global field-aligned mesh in the real space [22], and effective utilization of the most powerful supercomputers [23].

The current GTC simulations find that an electrostatic helically-trapped electron mode (HTEM) can be driven strongly unstable by a realistic density gradient in the W7-X high-mirror configuration. The linear eigenmode consists of many toroidal harmonies and is localized to discrete field lines on the inner side of the triangular section with weak magnetic field and unfavorable magnetic curvature, where the fraction of helically-trapped electrons is large. This HTEM can only be captured by global simulations since helically-trapped particles residing in different field lines can either drive or damp the HTEM. In contrast to earlier flux-tube simulations [11], global GTC simulations find that a strongly unstable HTEM can drive a large particle diffusivity comparable to the heat conductivity driven by the ITG turbulence with the same gradient scale lengths. The large particle transport driven by the HTEM turbulence could have significant implications on the confinement of fusion fuel tritium and the removal of fusion products Alpha ash in the fusion reactor.

*Global gyrokinetic simulation with kinetic electron.-* Global simulations of microturbulence with kinetic electron using the gyrokinetic toroidal code (GTC) have been extensively verified and validated for the microturbulence in tokamaks with axisymmetric magnetic fields [19-21] and with 3D resonant magnetic perturbations [24,25]. Recently, GTC has been applied to study ITG turbulence [26] and collisionless damping of zonal flows [27] in the LHD and W7-X stellarators, neoclassical and turbulent transport with self-consistent ambipolar electric fields in the W7-X [28], and ITG and

TEM with kinetic electrons in the LHD [15,29]. In the current electrostatic simulations, ions are governed by the nonlinear gyrokinetic equation [30]. All passing, toroidally-trapped, and helically-trapped particles are simulated using the drift-kinetic equation. Perturbed electrostatic potential is calculated using gyrokinetic Poisson equation [31].

The simulated W7-X equilibrium is from Ref. [32], which was calculated by the equilibrium code VMEC [33] and cast in the Boozer coordinates $(\psi, \theta, \zeta)$, where $\psi$ is poloidal flux, $\theta$ is poloidal angle, and $\zeta$ is toroidal angle [34]. This equilibrium has a small range of rotational transform $\iota$ =[0.88, 0.96], i.e., magnetic shear is nearly zero that violates the validity condition of the flux-tube simulation [35]. This configuration has a mirror-like magnetic field $B$ with the toroidal variation of $B$ much larger than the poloidal variation of $B$. Fig. 1a shows that the $B$ is largest at $\zeta=0$ with a bean-shaped cross-section and smallest at $\zeta=\pi/5$ with a triangular cross-section. Consequently, the amplitude of the helical component B(m,n)= B(1,5) is larger than the toroidal component B(1,0) [26], so there are more helically-trapped particles than the toroidally-trapped particles. For the HTEM simulation, the electron and ion temperatures are uniform with $T_i=T_e=1keV$ and a Maxwellian distribution. The electron density $n_e$ profile is defined by a gradient function $d \ln n_e / ds = -1 + |2s - 1|$ where $s = \psi/\psi_x$ and $\psi_x$ is the poloidal flux at the last closed flux surface. The ion is proton with a real mass ratio of $m_i/m_e$=1836. The major radius is $R_0$=5.61m and the on-axis magnetic field is $B_0$=2.43T at $\zeta = 0$ (left endpoint in Fig.1). The on-axis ion gyroradius is $\rho_i = 1.32 \times 10^{-3}$m. Since the W7-X has a field period of $N_f$=5, only one-fifth of the torus as shown in Fig. 1 is simulated with a periodic boundary condition in the toroidal direction. Thus, we only simulate the eigenmode family $n=kN_f+i$ with $i=0$, where $k$ is a positive number and $i$ indicates the different five eigenmode families. The radial simulation domain is $s = [0.2, 0.8]$, which is sufficiently large to incorporate nonlocal effects of magnetic drift of helically-trapped particles across magnetic surfaces on the turbulence dynamical time scale.

Full flux-surfaces with the entire poloidal angles are simulated to capture the true eigenmode structure in the non-axisymmetric stellarators with linear toroidal coupling of multiple toroidal harmonics $n$, which could result in the localization of eigenmodes to discrete magnetic field lines (equivalently, localization in the toroidal and poloidal angles). Global simulation would need an extremely large number of spatial grid points to resolve the HTEM eigenmode with the $n$ up to 350 (Fig. 4). Fortunately, this difficulty is overcome by the unique GTC capability using a global field-aligned mesh in the real space coordinates [22], which only needs a small number of parallel grid points because of the anisotropic eigenmode structures. Based on convergence studies [26], the current simulations use (121, 4400, 27) grid points, respectively, in the radial, poloidal, and parallel direction. In contrast, conventional global codes need 800 toroidal grid points, which is computationally prohibitive.

Numerical convergence studies show that nonlinear HTEM simulation requires 200 particles per cell for each species and a time step size for the ion orbit and field solver $\Delta t=0.008R_0/C_s$ where $C_s = \sqrt{T_e/m_i}$. The time step size for the electron guiding center pusher is 30 times smaller due to the small electron mass. The current simulation evolving the HTEM from linear to nonlinear steady state turbulence (Fig. 3) performs approximately $10^{15}$ particle-step calculations, which becomes feasible thanks to the excellent GPU-optimization of the GTC code in the most powerful supercomputers [23].

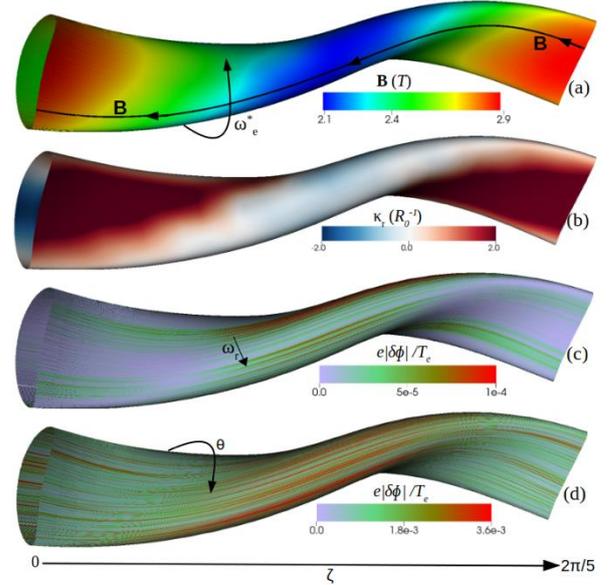

FIG.1. Flux surface at $s = 0.5$ is viewed from geometry center. (a) Magnetic field amplitude $B$. Black line is a magnetic field line on the inner side (θ~π), i.e., facing geometry center. (b) Magnetic curvature $\kappa_r$. (c) Electrostatic potential amplitude $e|\phi|/T_e$ in the linear phase at $t=16R_0/C_s$. (d) Electrostatic potential amplitude $e|\phi|/T_e$ in the nonlinear phase at $t=32R_0/C_s$.

*Linear excitation of HTEM.-* GTC global simulation of the W7-X with drift kinetic electrons using physical and numerical parameters described above finds an unstable driftwave excited solely by density gradients. The mode location coincides with the region of helically trapped electrons. This mode only exists with kinetic electrons since there is no instability in another simulation using adiabatic electrons. The linear eigenmode consists of many toroidal harmonics with dominant $n$=[150, 320] at $s$=0.5 (black line in Fig. 4). The maximal amplitude occurs at $n$=260, which corresponds to $k_\theta \rho_i = 1.24$. Here, poloidal wavevector $k_\theta$=950/m is calculated using the poloidal wavelength at $s$=0.5. The linear growth rate is $\gamma = 0.52\, C_S/R_0$ and the real frequency is $\omega_r = 1.40\, C_S/R_0$. It is excited by ion free energy associated with the density gradient, similar to that reported in Ref. [12]. It propagates in the ion diamagnetic direction and the frequency is much smaller than the diamagnetic frequency $\omega^* = k_\theta v_{dia} = 18.2$ (in the unit $C_S/R_0$). For comparison, ion acoustic wave frequency in the fluid limit is $\omega_{IAW} = 3.57$ with finite

Larmor radius (FLR) effects and $k_\parallel=7.59/R_0$ measured from the simulation. The ion transit frequency is $\omega_i^t = 7.59$ over a parallel wavelength and ion bounce frequency is $\omega_b^i = 3.20$ in the helical magnetic well. The ion guiding center magnetic drift frequency $\omega_i^d = k_\theta v_d = 0.856$ for the helically-trapped ion with thermal energy is in the same direction. The electron driftwave frequency is $\omega_e^* = -4.02$ with FLR effects and electron guiding center magnetic drift frequency $\omega_e^d = -0.856$ for the helically-trapped electron with thermal energy at $s=0.5$. Apparently, poloidal components of the guiding center magnetic drift and diamagnetic flow are in the same direction, so this W7-X equilibrium is not a maximum-J configuration for the helically-trapped particles (here J is the second invariant) [36,37].

Linear HTEM structures extend along the magnetic field lines and are localized in both toroidal and poloidal directions. The parallel wavelength is much longer than the perpendicular wavelength, as expected for the usual driftwave eigenmode. As shown in Fig. 1c, the linear eigenmode amplitude peaks in the weak magnetic field region (triangular section) with a toroidal angle $\zeta = \pi/5$ that has the largest fraction of helically-trapped particles. On the poloidal plane shown in Fig. 2a, the eigenmode amplitude peaks on the inner side of the triangular section with a poloidal angle $\theta \sim \pi$ (i.e., facing the geometry center), where the magnetic curvature points to the direction away from the geometry center and to the magnetic axis (i.e., bad curvature for normal pressure profile peaking on the magnetic axis). Fig. 1b shows the bad curvature on the inner side ($\theta \sim \pi$) of the triangular section at $\zeta=\pi/5$, but on outer side ($\theta \sim 0$) of the bean section at $\zeta=0$. Here, the radial component $\kappa_r$ of the magnetic curvature $\mathbf{\kappa} = (\mathbf{b} \cdot \nabla)\mathbf{b}$ is calculated as $\kappa_r = (\mathbf{\kappa} \cdot \nabla \psi)/|\nabla \psi|$ and $\mathbf{b}$ is the unit vector in the direction of the magnetic field.

Both the HTEM in the W7-X and the conventional TEM in the tokamak localize in the unstable region with weak magnetic field (where trapped particles reside) and bad curvature (which destabilizes driftwave), but the HTEM resides in the inner mid-plane of the W7-X while the TEM resides in the outer mid-plane of the tokamak [20]. More importantly, the TEM amplitude is uniform across all magnetic field lines, but the HTEM amplitude sharply peaks at magnetic field lines that center at the inner mid-plane of the triangular section of the W7-X. The localization of the HTEM structure to only a small fraction of the magnetic field lines shown in Fig. 1c is consistent with the coupling of many toroidal harmonics in the linear eigenmode spectrum shown in Fig. 4, which requires global simulations. In fact, the flux-tube simulations in Ref. [11] focus on the field lines centered at the outer mid-plane of the W7-X, and thus only find a weak TEM (but not the stronger HTEM).

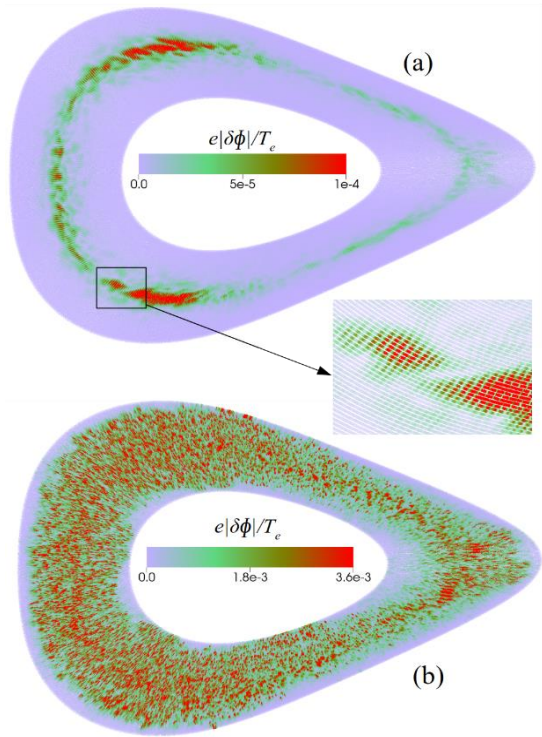

FIG.2. Electrostatic potential amplitude $e|\phi|/T_e$ on the poloidal plane with triangular cross-section at $\zeta=\pi/5$ for (a) linear phase at $t=16R_0/C_s$, and (b) nonlinear phase at $t=32R_0/C_s$. The left side is facing the geometry center. Color bars are the same as that in Fig. 1 (c) and (d).

*Nonlinear saturation of HTEM.-* After the initial exponential growth in the linear phase, the HTEM amplitude and particle diffusivity grow to a high level at $t \sim 20R_0/C$ and gradually saturate to reach a steady state turbulence after $t \sim 30R_0/C$ as shown by the red solid line in Fig.3. Similar turbulence amplitude and transport levels have been observed in another simulation where zonal flows [18] associated with flux-surface averaged potentials are artificially removed. Therefore, zonal flows are not the dominant saturation mechanism for the HTEM in the W7-X, similar to the TEM in the LHD stellarator [15]. On the other hand, zonal flows are the dominant saturation mechanism for the ITG in both the W7-X and LHD [26].

The saturation of the HTEM is accompanied by nonlinear processes of turbulence spreading in real space from unstable to stable regions and spectral transfer from unstable to damped modes. First, turbulence spreads in both radial and poloidal directions, eventually fills most of the linearly-stable radial domain and all poloidal angles as shown by the nonlinear mode structure in Fig. 1d and Fig. 2b. Nonetheless, the turbulence intensity is still the highest near the linear eigenmode structure. Secondly, the radially elongated streamers of the linear eigenmode are broken into more isotropic eddies due to the nonlinear ExB advection, which leads to the forward cascade of the radial wavevector spectrum to the linearly-damped, large radial wavevector regime. Finally, toroidal spectrum cascades inversely from linearly most unstable high-$n$ modes ($n\sim260$) to linearly-damped, but nonlinearly dominant, mesoscale-$n$ modes ($n\sim100$) as shown in Fig.4. The HTEM saturates when the nonlinear transfers of the energy from waves to particles balances the linear drive by the density gradients.

Although zonal flows do not dominate the HTEM saturation directly, they can facilitate the inverse cascade of the toroidal spectrum because of the linear coupling between zonal modes (i.e., $n = 0$) and non-zonal modes (i.e., $n > 0$) in the stellarator [27]. At the early nonlinear phase $t=20R_0/C_s$ (green line in Fig. 4), linearly-stable toroidal harmonics ($n<100$) start to be generated with the $n=5$ harmonic of the perturbed electrostatic potential having the largest amplitude since the largest component of the non-axisymmetric equilibrium magnetic field is the $n=5$ harmonic. The amplitudes of these low-$n$ harmonics increase quickly during the saturation process (pink line for $t=24R_0/C_s$). The high-$n$ unstable harmonics can scatter on these low-$n$ harmonics, leading to the inverse cascade of the toroidal spectrum [38]. After the saturation, the toroidal spectrum of steady state turbulence at $t=32R_0/C_s$ (red line) is dominated by the mesoscale-$n$ harmonics ($n\sim100$).

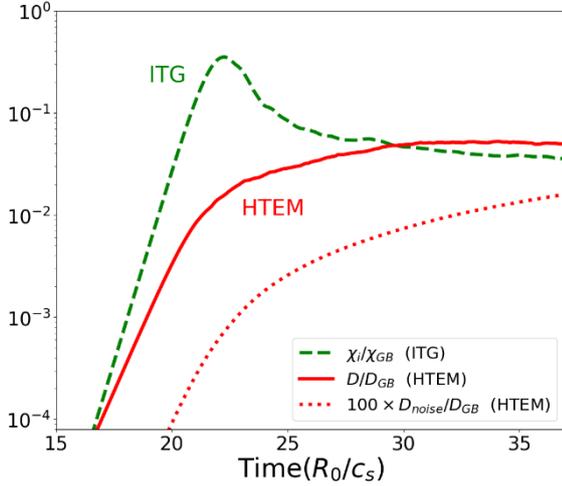

FIG.3. Time evolution of flux-surface averaged particle diffusivity (solid red) and noise-driven diffusivity (dotted red) in the HTEM simulation, and ion heat conductivity (dashed green) in the ITG simulation. Transport coefficients are calculated at the s=0.5 flux-surface and normalized by gyro-Bohm unit $D_{GB}$.

*HTEM transport.-* In the steady state turbulence, the HTEM drives a particle diffusivity (same for both electron and ion because of quasi-neutrality) $D \approx 0.08 D_{GB}$, where $D_{GB} = \rho^* c T_e/eB$ is the gyroBohm unit and $\rho^* = \rho_i/L_n$ is the ion gyroradius normalized by the density gradient scale length $L_n$ with $R_0/L_n = 21.3$ at the $s = 0.5$ flux-surface. To make sure that the simulation is not contaminated by particle noise, the noise-driven diffusivity [39] is also calculated in the same simulation and shown in Fig. 3. The noise-drive diffusivity represents less than 1% of the total measured particle diffusivity, which indicates that the nonlinear simulation is numerically converged.

The effectiveness of the HTEM turbulence in driving the particle transport is comparable to that of the ITG turbulence in driving the heat transport. For comparison, we have performed another GTC simulation with kinetic electrons using the same W7-X geometry and all parameters, except that the density is now uniform and that the electron temperature (same as ion temperature) is defined by the following gradient function: $d \ln T_e/ds = -1 + |2s - 1|$, which is the same as the density gradient function in the HTEM simulation. The new simulation with this temperature profile finds an ITG instability with a linear growth rate $\gamma = 0.86\, C_S/R_0$ and a real frequency $\omega_r = 3.22\, C_S/R_0$. The ITG nonlinear saturation is dominated by the zonal flows [26]. The steady state ITG turbulence drives an ion heat conductivity (green dashed line in Fig. 3) very similar to the particle diffusivity driven by the HTEM turbulence. Therefore, global GTC simulation results finding a large particle transport driven by the strong HTEM turbulence contradict earlier studies [11] predicting a benign TEM in the W7-X. Finally, GTC simulations with both density and temperature gradients have also been performed and confirmed earlier finding [40,6] that the transport is lower with both density and temperature gradients. More detailed study will be reported in a future publication.

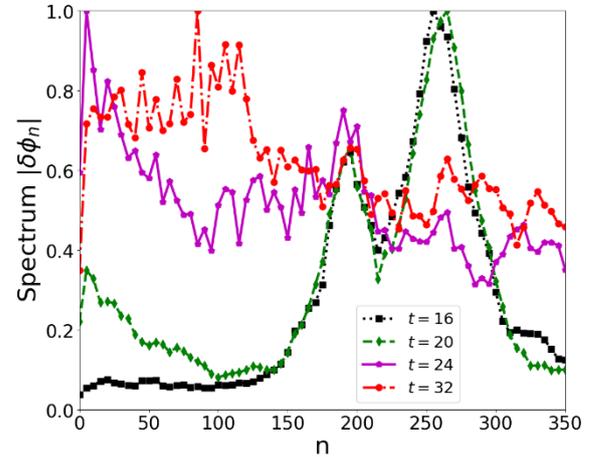

FIG.4. Time evolution for toroidal spectrum of perturbed electrostatic potential at s=0.5 flux-surface from linear phase ($t=16R_0/C_s$), to nonlinear saturation ($t=20$-$24R_0/C_s$), and finally to steady-state turbulence ($t=24R_0/C_s$).

The W7-X model equilibrium used in this work was originally generated by J. Geiger. The authors would like to thank J. Riemann, R. Kleiber, and D.A. Spong for providing the equilibrium data. This work was supported by the U.S. DOE grant DE-FG02-07ER54916 and INCITE program. Simulations used the computing resources at ORNL (DOE Contract DE-AC05-00OR22725) and NERSC (DOE Contract DE-AC02-05CH11231).

**References**
[1] H. E. Mynick *et al*, Phys. Rev. Lett. **48**, 322 (1982).
[2] A. H. Boozer, Phys. Plasmas **5**, 1647 (1998).
[3] D. A. Spong, Phys. Plasmas **22**, 055602 (2015).
[4] A. Dinklage *et al*, Nature Phys. **14**, 855 (2018).
[5] C. D. Beidler *et al*, Nature **596,**221(2021).
[6] W. M. Tang, Nucl. Fusion **18**, 1089 (1978).
[7] N. A. Pablant *et al*, Phys. Plasmas **25**, 022508 (2018).

[8] T. Klinger *et al*, Nucl. Fusion **59**, 112004 (2019).
[9] E. M. Edlund *et al*, Rev. Sci. Instrum. **89**, 10E105 (2018).
[10] J. H. E. Proll *et al*, Phys. Rev. Lett. **108**, 245002 (2012).
[11] J. H. E. Proll *et al*, Phys. Plasmas **20**, 122506 (2013).
[12] G. G. Plunk *et al*, J. Plasma Phys. **83**, 715830404 (2017).
[13] E. Sánchez *et al*, Nucl. Fusion **61**, 116074 (2021).
[14] J. Riemann *et al*, Plasma Phys. Control. Fusion **64**, 104004 (2022).
[15] T. Singh *et al*, Nucl. Fusion **62**, 126006 (2022).
[16] E. Sánchez *et al*, Nucl. Fusion **63**, 046013 (2023).
[17] A. Mishchenko *et al*, J. Plasma Phys. **89**, 955890304 (2023).
[18] Lin Z. Lin *et al*, Science **281**, 1835 (1998).
[19] Z. Lin *et al*, Plasma Phys. Control. Fusion **49**, B163 (2007).
[20] Y. Xiao and Z. Lin, Phys. Rev. Lett. **103**, 085004 (2009).
[21] P. Liu *et al*, Phys. Rev. Lett. **128**, 185001 (2022).
[22] Z. Lin *et al*, Phys. Rev. Lett. **88**, 195004 (2002).
[23] W. Zhang *et al*, Lecture Notes in Computer Science **11381**, 3 (2019).
[24] I. Holod *et al*, Nucl. Fusion **57**, 016006 (2016).
[25] K. S. Fang and Z. Lin Phys. Plasmas **26**, 052510 (2019).
[26] H. Y. Wang *et al*, Phys. Plasmas **27**, 082305 (2020).
[27] J. H. Nicolau *et al*, Nucl. Fusion **61**, 126041 (2021).
[28] J. Y. Fu *et al*, Phys. Plasmas **28**, 062309 (2021).
[29] T. Singh *et al*, submitted to Nucl. Fusion (2023).
[30] A. J. Brizard and T. S. Hahm Rev. Mod. Phys. **79**, 421 (2007).
[31] W. W. Lee, Phys. Fluids **26**, 556 (1983).
[32] J. Riemann *et al*, Phys. Plasmas Control. Fusion **58**, 074001 (2016).
[33] S. P. Hirshman and J. C. Whitson, Phys. Fluids **26**, 3553 (1983).
[34] A. H. Boozer, Phys. Fluids **24**, 1999 (1981).
[35] M. A. Beer *et al*, Phys. Plasmas **2**, 2687 (1995).
[36] P. Helander *et al*, Phys. Plasmas **20**, 122505 (2013).
[37] P. Helander, Rep. Prog. Phys. **77**, 087001 (2014).
[38] Z. Lin *et al*, Phys. Plasmas **12**, 056125 (2005).
[39] I. Holod and Z. Lin, Phys. Plasmas **14**, 032306 (2007).
[40] J. A. Alcusón *et al*, Plasma Phys. Control. Fusion **62**, 035005 (2020).